\documentclass[aps,prd,twocolumn,showpacs,superscriptaddress,groupedaddress]{revtex4-1}

\usepackage{graphicx}  
\usepackage{dcolumn}   
\usepackage{bm}        
\usepackage{amssymb}   
\usepackage{amsmath}   

\usepackage{color} 

\usepackage{multirow}
\usepackage{enumerate}
\usepackage{comment}
\usepackage{tabularx}
\usepackage{booktabs}
\usepackage{float}
\usepackage{gensymb}
\usepackage{siunitx}

\begin{document}

\title{Newtonian-noise reassessment for the Virgo gravitational-wave observatory including local recess structures}

\author{Ayatri Singha}
\email{a.singha@maastrichtuniversity.nl}
\affiliation{Department of Gravitational Waves and Fundamental Physics, Maastricht University, P.O. Box 616, 6200 MD Maastricht}
\affiliation{Nikhef, Science Park 105, 1098 XG Amsterdam, The Netherlands}

\author{Jan Harms}%
\email{jan.harms@gssi.it}
\affiliation{Gran Sasso Science Institute (GSSI), I-67100 L'Aquila, Italy}
\affiliation{INFN, Laboratori Nazionali del Gran Sasso, I-67100 Assergi, Italy}
\author{Stefan Hild}
\email{stefan.hild@maastrichtuniversity.nl}
\affiliation{Department of Gravitational Waves and Fundamental Physics, Maastricht University, P.O. Box 616, 6200 MD Maastricht}
\affiliation{Nikhef, Science Park 105, 1098 XG Amsterdam, The Netherlands}
\date{\today}

\begin{abstract}
The LIGO and Virgo scientific collaborations have cataloged ten confident detections from binary black holes and one from binary neutron stars in their first two observing runs, which has already brought up an immense desire among the scientists to study the universe and to extend the knowledge of astrophysics from these compact objects. One of the fundamental noise sources limiting the achievable detector bandwidth is given by Newtonian noise arising from terrestrial gravity fluctuations. It is important to model Newtonian noise spectra very accurately as it cannot be monitored directly using current technology. In this article, we show the reduction in the Newtonian noise curve obtained by more accurately modelling the current configuration of the Virgo observatory. In Virgo, there are clean rooms or recess like structures underneath each test mirror forming the main two Fabry-Perot arm cavities of the detector. We compute the displacements originating from an isotropic Rayleigh field including the recess structure.  We find an overall strain noise reduction factor of 2 in the frequency band from 12 to about 15 Hz relative to previous models. The reduction factor depends on frequency and also varies between individual test masses.
\end{abstract}

\maketitle

\section{Introduction}

Currently operating broadband gravitational-wave detectors have opened up a new window to study the universe \cite{AbEA2016a,AbEA2017d,MM170817,PhysRevX903104}. Detector sensitivity in the low-frequency range plays a crucial role to accumulate more information for the inspiral parameters of the gravitational-wave sources \cite{LyEA2015,HaHi2018}. In this frequency range (below 20\,Hz), detector sensitivity is at least partly limited by Newtonian noise (NN) or gravity-gradient noise \cite{Sau1984,Har2019}. Newtonian noise arises from the fluctuation of local gravitational fields. The test mirrors in the interferometer are subject to gravity perturbations due to the propagation of seismic waves, atmospheric changes, etc. It is important to model the NN spectrum accurately as so far we could not measure this instrumentally. 

In order to model the NN curve, we need to monitor the local seismic field nearby the input and output test mirrors (can be obtained from seismometer and tiltmeter data deployed near the test mirrors). In general, seismic waves can be categorized into surface waves (Rayleigh and Love waves) and body waves (P and S waves) \cite{AkRi2009}. Rayleigh waves can propagate on surfaces of homogeneous media, and their magnitude decays exponentially with depth. Love waves are surface shear waves guided in near surface layered media. Love waves are not of concern here since they do not produce NN \cite{Har2019}. The body waves, comprised of compressional and shear waves, can propagate through media in all directions. 

For the LIGO and Virgo detectors, the dominant seismic sources are part of the detector infrastructure (pumps, ventilation,...), and produce predominantly surface waves, which means that our subsequent analysis can focus on NN from Rayleigh waves \cite{HuTh1998,BeEA1998,CoEA2016a, CoEA2018a}. Seismic NN is estimated to be the main contribution to the overall NN compared to other contributions like acoustic NN or NN from infrastructure \cite{DHA2012}. Although the effect of NN cannot be screened out directly from the detector, still there are some conventional ways to reduce this noise. First, one can select a seismically quiet region to build up the detector \cite{HaEA2010,BBR2015,CoEA2014,CoEA2018b}. For example, the proposed next-generation European detector Einstein Telescope is planned to be built underground to minimise the Newtonian noise \cite{ET2011,BaHa2019}. Another method to mitigate NN that can also be applied to current detectors is by employing seismic sensors to use their data for a coherent cancellation of NN \cite{Cel2000,HaVe2016,CoEA2018a}. 

In 2014, Harms and Hild have shown that by changing the local topography surrounding the test mass, it is also possible to reduce NN for surface detectors \cite{HaHi2014}. Without a recess, the linear (root power) NN spectral density of test-mass displacement assuming a flat surface can be estimated as
\begin{equation}
X_{\rm NN}(f) = \frac{1}{\sqrt{2}}2 \pi \gamma G \rho_0 \frac{\xi(f)}{(2\pi f)^2} \exp(- 2 \pi h/\lambda),
\end{equation}
where $\xi(f)$ is the linear spectral density of vertical ground vibration assumed to be dominated by an isotropic Rayleigh-wave field, $G$ is Newton's constant, $\rho_0$ is the density of the surface medium, $h$ is the height of the test mass above ground, and $\lambda$ the length of a Rayleigh wave at frequency $f$. 

A recess structure can effectively lower the density close to the test mass, or in other words, increase the height $h$ of a test mass above ground. This, in turn, results in the reduction of the NN spectra in a certain frequency band determined by the size of the recess structure and to some extent also on the precise shape. Harms and Hild have also calculated how the NN reduction factor varies if one adjusts the recess parameters showing that feasible constructions could lead to significant NN reduction by factors 2 -- 4. However, for currently operating detectors, one will hardly be eager to modify the experimental setup by removing ground under the mirrors and setting up pillars to support the vacuum chambers. However, the Virgo detector was constructed with recess structures, cleanrooms under test masses used for test-mass installation from below the vacuum chambers, which leads to the question to what extent NN is reduced. Previously, Virgo NN spectra were calculated using analytical equations for a flat surface. In this paper, we evaluate the NN spectra numerically including all the dimensions of the clean rooms in the central, north and west-end buildings of Virgo. We present the reduction factor in the overall NN strain spectra for the isotropic Rayleigh-wave field and present an analysis how it depends on direction of wave propagation.

In section  \ref{Recess_parameters}, we sketch the proper dimension of the recesses under the input and end test mirrors inside central and end buildings of Virgo. In section \ref{Recess_factors}, we simulate the isotropic Rayleigh-wave field in the vicinity of the test mirrors and estimate the gravity perturbation on each test mirror using a finite-element model. We show the reduction in gravity perturbation on the test mirror due the presence of the recess. In section \ref{NN_Ref}, we compute the reduction in NN for each test mirror in Virgo and combine them to get the overall reduction factors. Additionally, in section \ref{sec:directional}, we consider the directional seismic field and analyze how the NN reduction factors (for end test masses in Virgo) vary with the direction of propagation of the seismic field. 

\section{Recess parameters}
\label{Recess_parameters}
\begin{figure*}[ht!]
    \includegraphics[width=\textwidth]{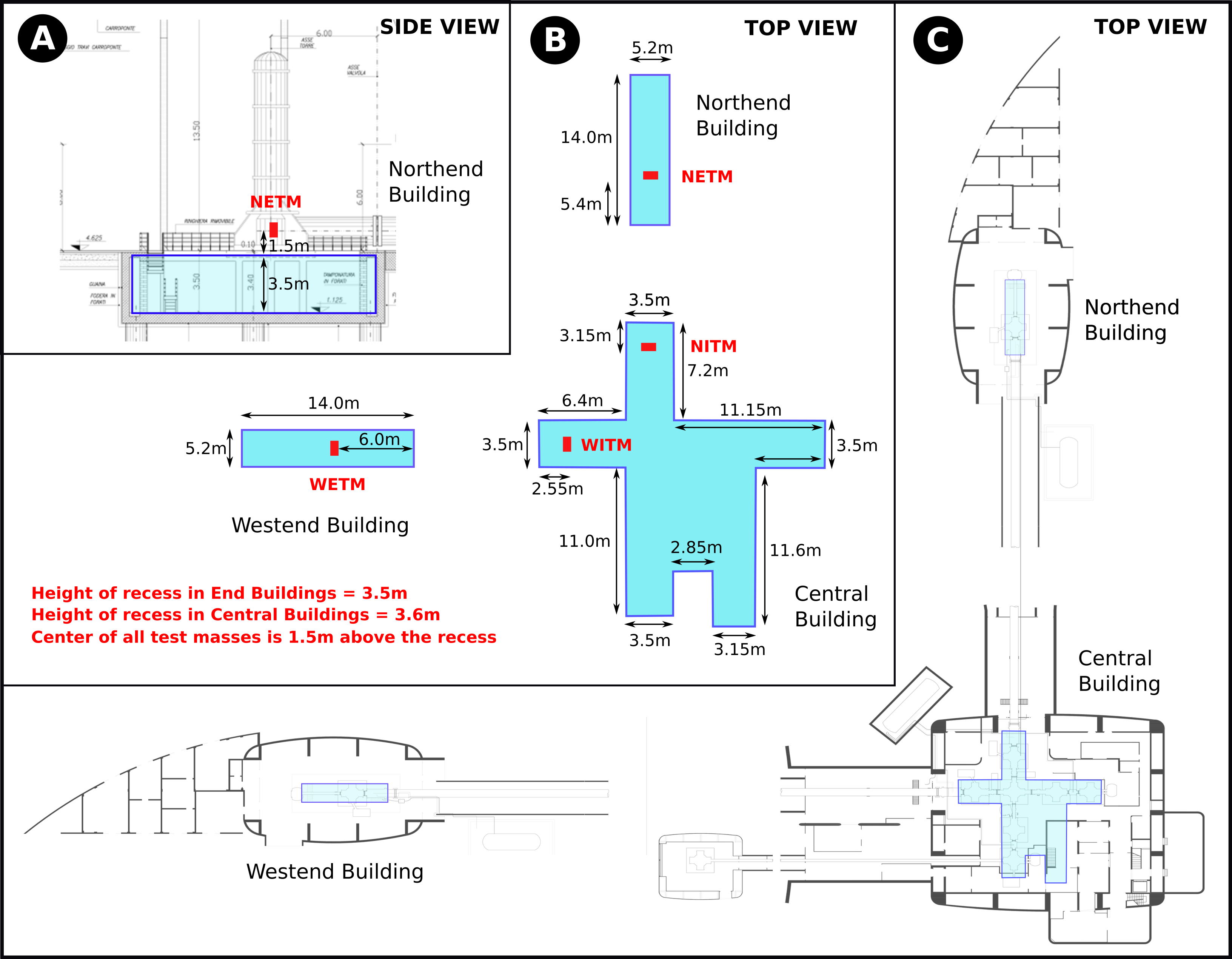}
    
\caption{Overview of the Virgo configuration(with recess parameters bellow test mirrors). The sky-blue shadow in this figure represents the position of the clean-rooms or the recesses under the test mirrors. Part (B): TOP VIEW, the red dots are the test mirrors. Part (A): SIDE VIEW, we can see the location of such recess in the ground label. The depth of the recesses in the end buildings is 3.5\,m and for the central building, it is about 3.6\,m. For each case, the mirror is hanging 1.5\,m above surface level.}
\label{fig:Recess_dim}
\end{figure*}

The goal of this study is to recalculate the NN spectrum taking into account the clean rooms underneath the test mirrors in the observatory buildings. Hence, this modification is only valid for Virgo as we are considering the specific dimension of the space surrounding the test mirrors. In this section, we show all the dimensions of the recess in the Central Building (CEB), North End Building (NEB) and West End Building (WEB). Figure \ref{fig:Recess_dim} shows schematic geometries of the recesses. The test mirrors are at a height of 1.5\,m above surface level. The shape of the recess for end test mirrors (ETMs) is a simple rectangle, but not symmetric along each arm. For the input test mirrors (ITMs), we have considered the whole clean-room space in the central building, which is much more extended and less symmetrical.

\section{Suppression factors of gravity perturbation on test mirrors due to presence of recesses}
\label{Recess_factors}
\begin{figure}[ht!]
    \includegraphics[width=\columnwidth]{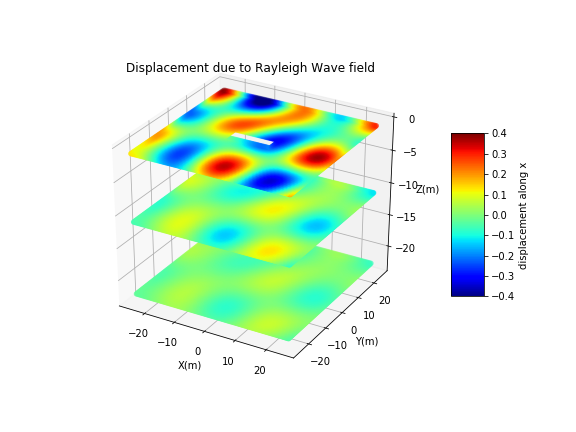}
    \includegraphics[width=\columnwidth]{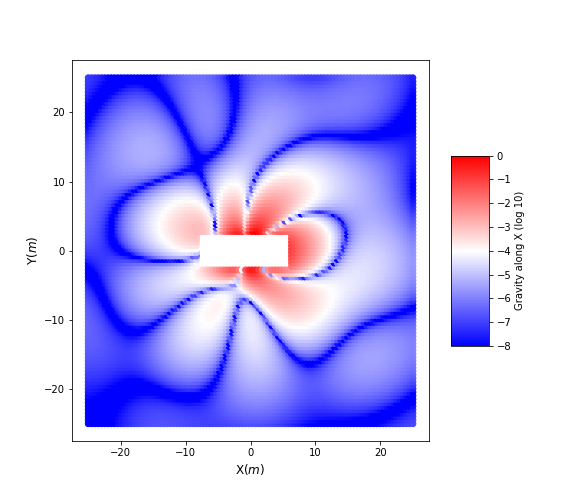}
\caption{(top) Displacements due to the Rayleigh-wave field for a few surface layers. The amplitude of the field decreases with depth. (bottom) Contribution to the Rayleigh gravity perturbation from different parts of the surface displacement field. The layers have been shown including the recess.}
\label{fig:wavefield}
\end{figure}

We consider the isotropic Rayleigh-wave field for various frequencies from 5 to 25\,Hz. We generate the isotropic field in our simulation by adding subsequently the contribution of plane Rayleigh waves propagating along random directions. The detailed mathematical formulation for the Rayleigh waves and its intrinsic parameters can be obtained in \cite{HaEA2009a}. In  figure \ref{fig:wavefield} (top), we show the vertical displacement field surrounding a test mirror for different surface layers. The color bar indicates the magnitude of the Rayleigh field at each grid point. The amplitude of the Rayleigh field decays exponentially with depth. In this plot, the shape of the recess is much smaller than the Rayleigh wavelength, which means that seismic scattering can be neglected.

We estimate the gravity perturbation due to density fluctuations from propagating seismic waves. A discrete version of the integral over particle displacement giving rise to gravity perturbations at the position of the test mirror ($r_0$) can be written as
\begin{equation}
     \delta a(r_0, t) = G\rho_0 \sum_{i}V_i \frac{1}{|\vec{r}_i - \vec{r_0}|^3}\bigg(\vec{\xi}(\vec{r}_i, t) - 3(\vec{e}_i\cdot \vec{\xi}(\vec{r}_i, t))\cdot\vec{e}_i\bigg),
    \label{eq:dipoleNN}
\end{equation}

where $\vec{r}_i $ denotes the position of i'th grid point, $\vec{\xi}(\vec{r}_i, t)$ is the corresponding Rayleigh displacement, and $\vec{e}_i $ is the unit vector pointing to $\vec{r}_i$ from $\vec{r}_0$ ($\vec{e}_i = \frac{\vec{r}_i - \vec{r}_0}{|\vec{r}_i - \vec{r}_0|} $). Summing over the finite-element model, we take into account gravity perturbations from vertical surface displacement as well as due to (de)compression of rock below surface.
 
We use this equation to evaluate the gravity perturbation based on a finite-element model of the ground. The displacement of the points from their equilibrium position produces a fluctuating gravitational field on the test mirror resulting in NN. Figure \ref{fig:wavefield} (bottom) shows the component of the gravity perturbation (along the direction of the arm) on the test mirror due to an isotropic Rayleigh field. 
\begin{figure}[ht!]
    \includegraphics[width=\columnwidth]{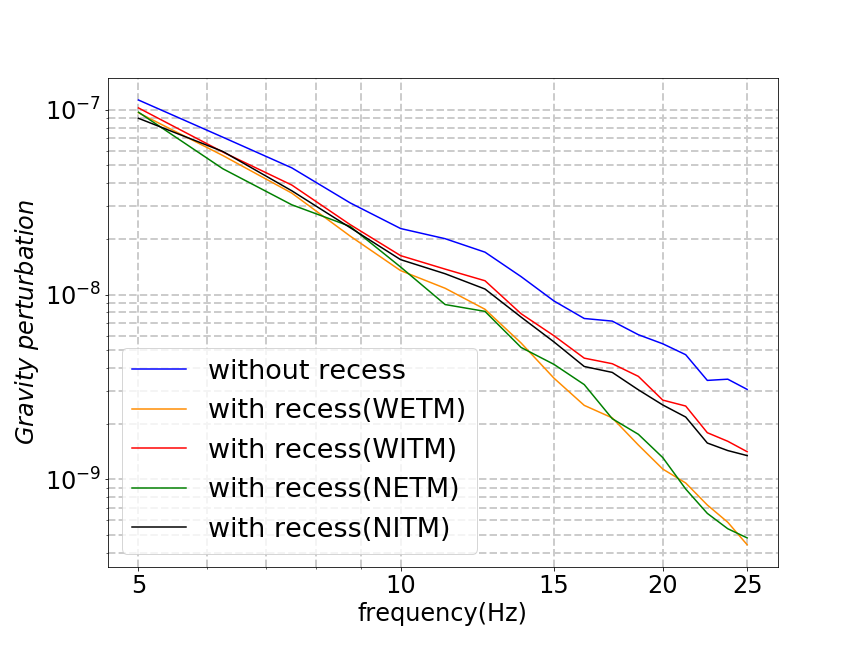}
    \includegraphics[width=\columnwidth]{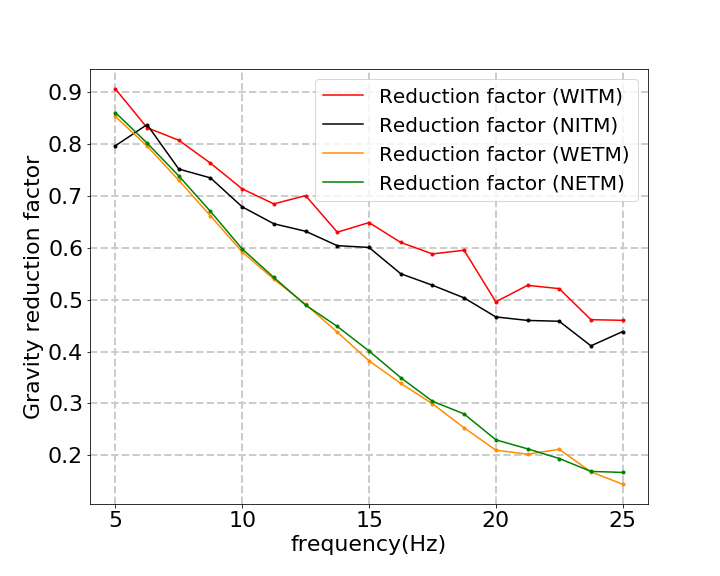}
\caption{(top) Gravity perturbation due to the propagation of Rayleigh waves on input and end test mirrors (ITM and ETM) of the west (W) and north (N) arm. (bottom) Reduction ratio with recess vs without a recess.}
\label{fig:recess_compare}
\end{figure}

We generate the Rayleigh field with uniform random propagation directions for each reference frequency between 5 and 25\,Hz and then compute the gravity perturbation on the test mirror. We replicate this calculation for two cases. First, we consider a symmetric grid structure surrounding the test mirror without having any recess. Hence, the gravity perturbation without a recess should be the same for each individual test mirror. While in the second case, we evaluate the gravity perturbation extracting the part of recess from the finite-element model. We repeat this calculation for each test mirror as the dimension (length and width) of the recess is different. 

In figure \ref{fig:recess_compare} (top), we show the gravity perturbation on test masses due to the isotropic seismic field with and without a recess. As expected, the magnitude of gravity perturbation is lowered by considering the recess structure, and the effect is more pronounced for ETMs compared to ITMs. In figure \ref{fig:recess_compare} (bottom), we plot the reduction ratio of gravity perturbation, which is simply the ratio of the perturbation with and without a recess. For ETMs we obtain a significant reduction (by a factor 2) for the frequency range between 10 to 15$\,$Hz. 

\begin{figure*}[ht!]
  \centering
    \includegraphics[width=\columnwidth]{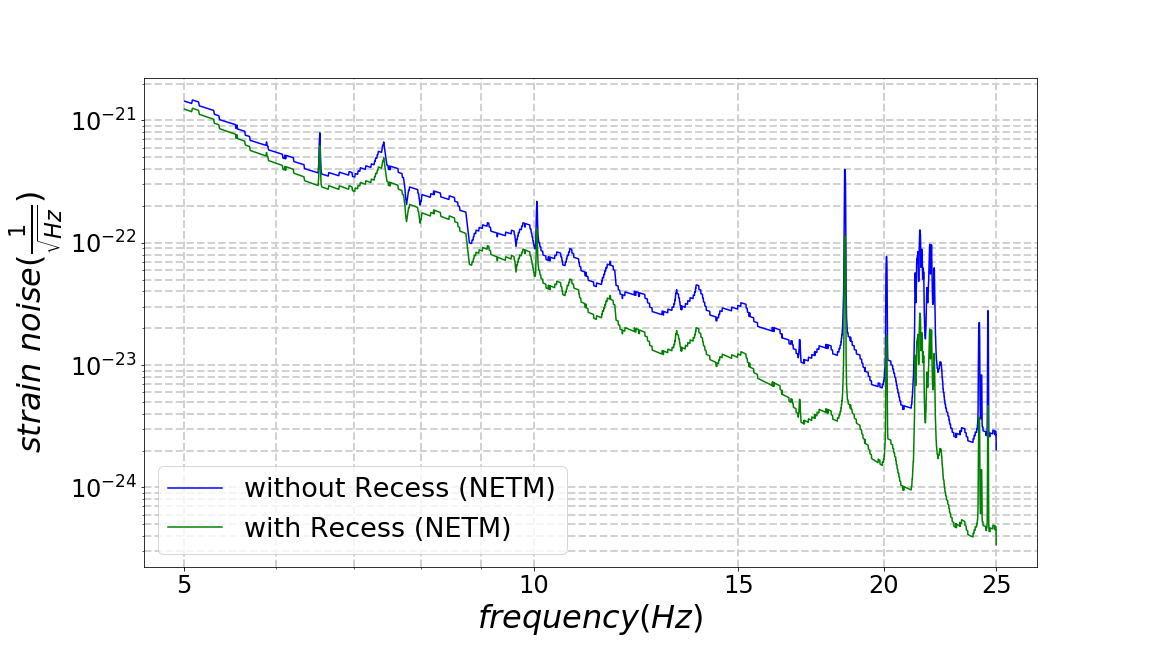}
    \includegraphics[width=\columnwidth]{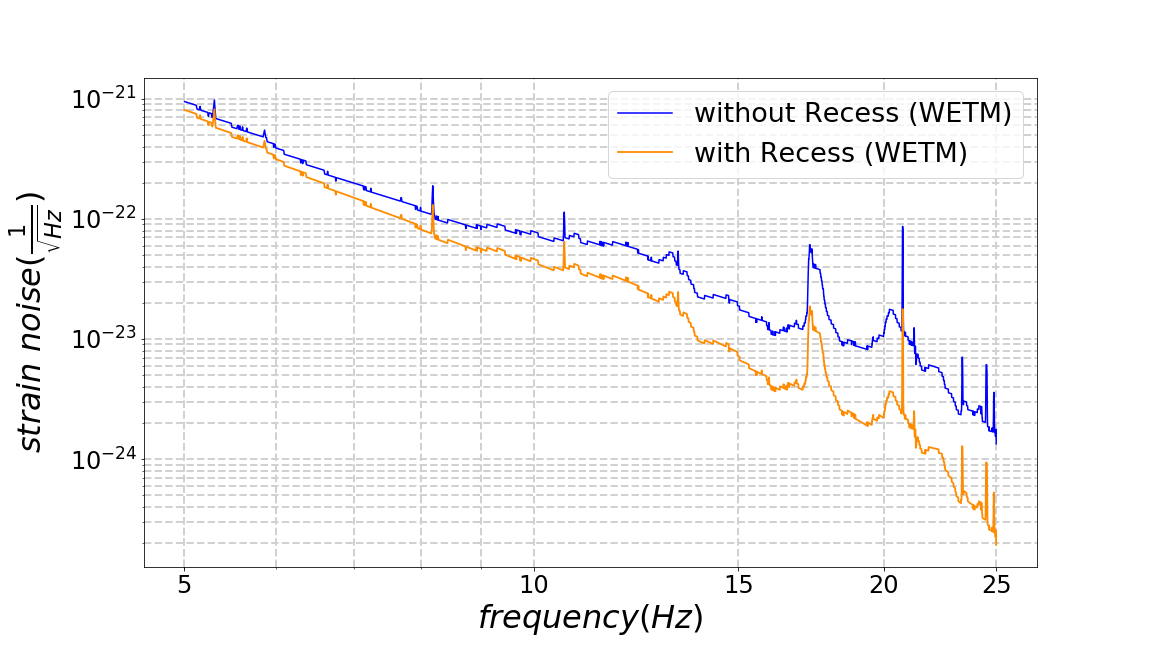}
    \includegraphics[width=\columnwidth]{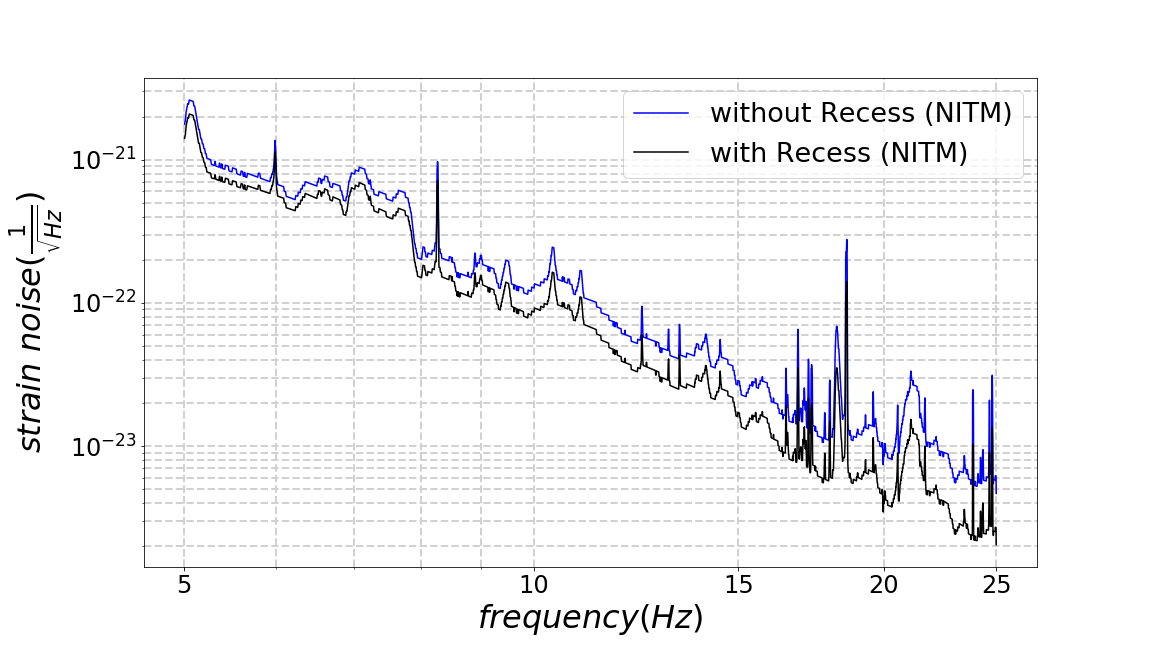}
    \includegraphics[width=\columnwidth]{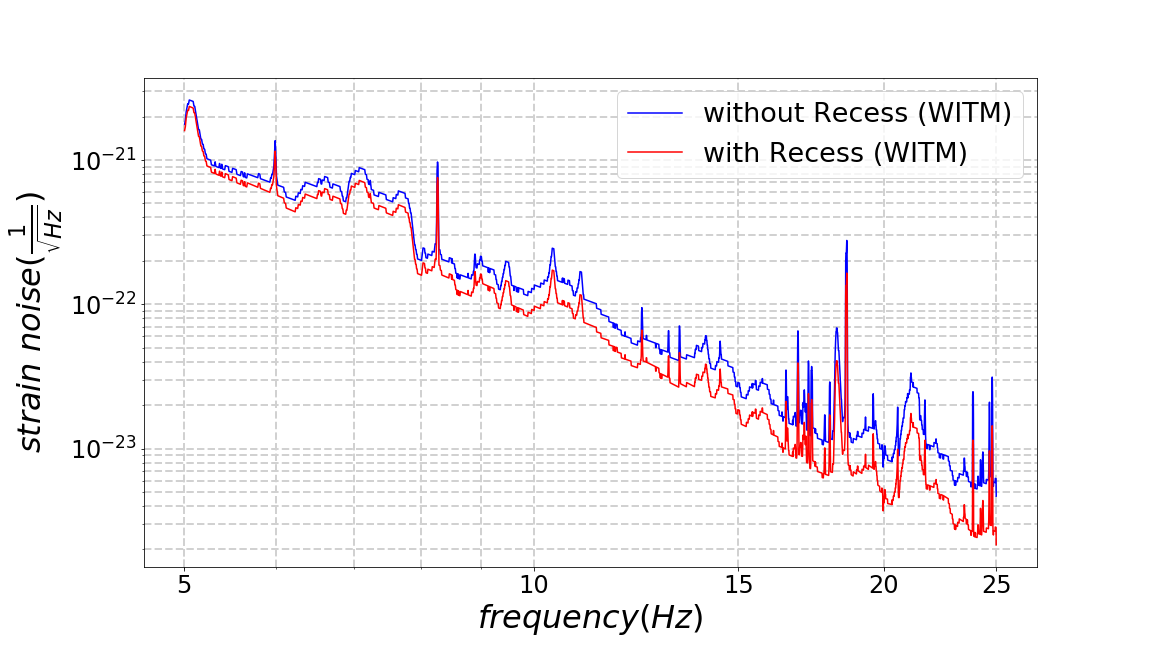}
\caption{(upper left) Newtonian noise spectra for North end test mirror (horizontal displacement along North). (upper right) Newtonian noise spectra for West end test mirror (horizontal displacement along West). (lower left) Newtonian noise spectra for North input test mass (horizontal displacement of NITM along north). (lower right) Newtonian noise spectra for West input test mass (horizontal displacement of WITM along West). For each sub-plots, the blue trace represents the spectra without a recess and the other spectra is associated with the corresponding mirror.}
\label{fig:NoiseSpectraITM}
\end{figure*}
The results show that NN reduction due to a recess is more significant at higher frequencies. This can be explained by the fact that the recess dimension must be compared with the length of Rayleigh waves. At frequencies where Rayleigh waves are much larger than the recess, the recess only excludes a minor part of all the relevant density fluctuations in the ground produced by these waves. The situation improves with shorter wavelength with the caveat that when Rayleigh waves become very short (in Virgo, at frequencies $>$15\,Hz \cite{TrEA2019}), scattering from the recess becomes important possibly leading to modifications of NN reduction not described by our model. 

\section{Result of Newtonian noise suppression of Virgo}
\label{NN_Ref}
In this section, we estimate the reduction in the absolute level of NN due to the presence of the recess underneath the test mirrors. Figure \ref{fig:NoiseSpectraITM} shows the NN spectra where we consider the seismic spectra recorded from seismometers deployed at the central building, and north- and west-end buildings of Virgo. We have used the 90th percentiles of vertical seismic spectra to normalize the vertical surface displacement in the finite-element model. The model itself then produces vertical and horizontal displacements at the surface and underground consistently according to an analytical equation of the Rayleigh-wave field. We have already found the reduction factors numerically for each test mirror. Just by multiplying those reduction factors for each test mirror with a no-recess NN estimate, we obtain the NN spectra with recess shown in figure \ref{fig:NoiseSpectraITM}. The NN spectra without a recess (blue curves) are added for comparison.

\begin{figure}[ht!]
  \centering
    \includegraphics[width=\columnwidth]{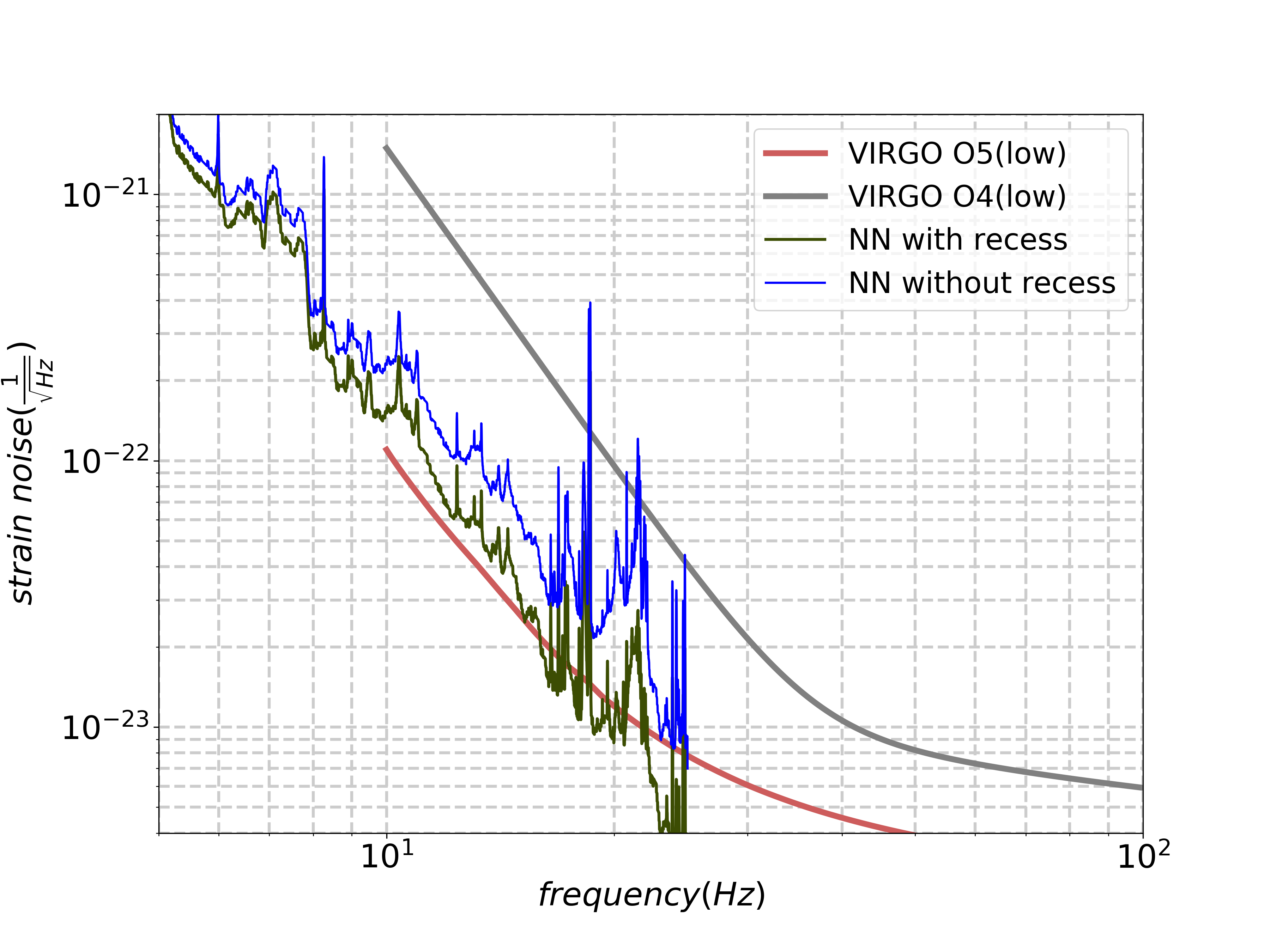}
    \includegraphics[width=\columnwidth]{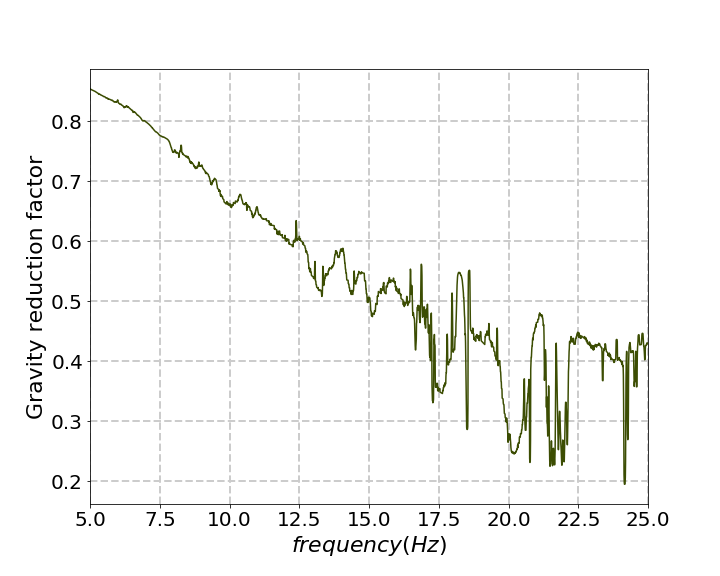}
\caption{(top) Total NN spectrum and its reduction when considering the recesses under test masses. For reference, we have also included the sensitivity of Virgo O4 and O5 for their lower limit. (bottom) Newtonian noise reduction ratio. Newtonian Noise suppression at frequencies above 15\,Hz in these plots may be significantly affected by seismic scattering (as one can estimate from analytical expressions, which are consistent with observations at the Virgo site \cite{TrEA2019}), which is not considered in the numerical simulation used for this analysis.}
\label{fig:Newtonian_noise_reduction}
\end{figure}

Finally, we estimate the overall NN curve shown in figure \ref{fig:Newtonian_noise_reduction} (top). Assuming that NN is uncorrelated between test masses (certainly valid for correlations involving ETMs due to their large distance to other test masses, but also valid for the ITM pair when assuming an isotropic seismic field), NN power spectral densities from individual mirrors can simply be added, and we then plot the square root of the total power spectral density. In figure \ref{fig:Newtonian_noise_reduction} (bottom), we show the reduction ratio for the overall NN spectra. We find the reduction in NN nearly a factor of 2 within 12 to 15\,Hz. It is not a smooth curve anymore (compare with figure \ref{fig:recess_compare}) since reduction factors for each test mass are different, which means that the ratio of total NN with and without a recess now depends on the shape of the individual spectra.  

The top plot in figure \ref{fig:Newtonian_noise_reduction} shows how important it is to model the effect of the recess. It has a significant impact on how much additional NN cancellation needs to be achieved with seismometer arrays.

\begin{figure}[ht!]
  \centering
    \includegraphics[width=\columnwidth]{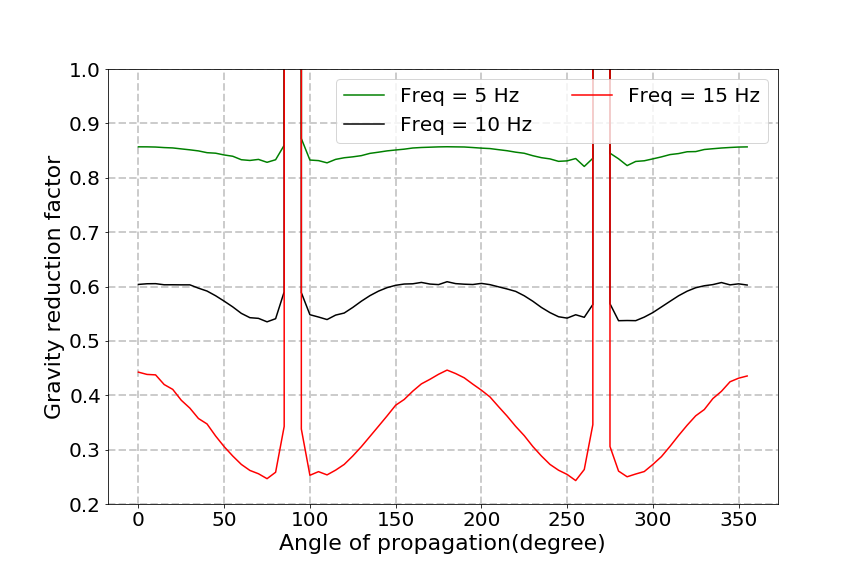}
    \includegraphics[width=\columnwidth]{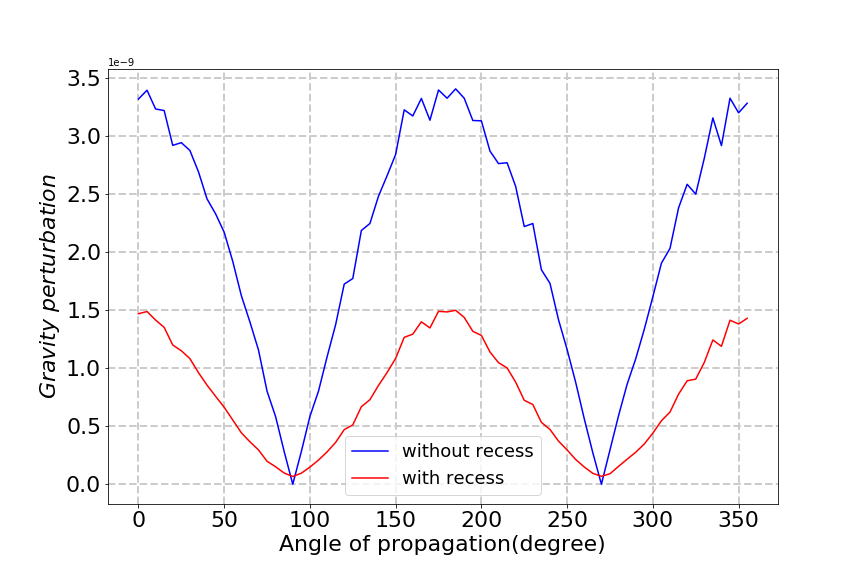}
\caption{(top) Reduction factor vs the direction of propagation of Rayleigh waves for reference frequencies 5, 10 and 15\,Hz. (bottom) Gravity perturbation with respect to angle of propagation for reference frequency 15\,Hz. }
\label{fig:Newtonian_noise_reduction_variation}
\end{figure}

\section{Newtonian-noise reduction due to recess in a directional seismic field}
\label{sec:directional}
In the previous section, we have seen the suppression of NN when one considers the presence of recesses bellow each test mirror. We have assumed the seismic field to be isotropic, i.e., Rayleigh waves incident from all directions with equal probability. However, it is likely that local seismic sources produce a significant anisotropy, which is why the impact of a recess needs to be understood as a function of propagation direction to be able to estimate corrections to the isotropic results.

As we can see, the shape of the recess is not spherically symmetric, more like a rectangle for end test mirrors (ETMs). So, we expect the reduction factor should have a dependence on the angle of propagation $\theta$ of a seismic wave. We evaluate this by computing the NN reduction factor for individual Rayleigh waves as a function of propagation direction. In figure \ref{fig:Newtonian_noise_reduction_variation} (top), we show how the reduction factor for the NN of the West ETM varies when the seismic field approaches from an angle $\theta$ ($0^\circ \leq \theta \leq 360^\circ$) with respect to the direction of the arm. Here we choose the values for three reference frequencies say, 5, 10 and 15\,Hz. In figure  \ref{fig:Newtonian_noise_reduction_variation} (bottom), we show the absolute gravity perturbation (in arbitrary units) with and without a recess for the reference frequency 15\,Hz. When the wave arrives along directions perpendicular to the arm ( $\theta = 90^\circ, 270^\circ$), the density fluctuation along the direction of the arm is zero without any recess. As the recess surrounding the mirrors is not symmetric, we see some unbalanced gravity field when considering recess, which means that NN never fully vanishes. Hence, when we are taking the ratio of the gravity perturbation with and without a recess, we are finding diverging factors for $\theta = 90^\circ, 270^\circ$. This should not be the case if the recess would be symmetric about the test mirror. The results in figure \ref{fig:Newtonian_noise_reduction_variation} can be used to correct NN spectra as shown in figure \ref{fig:Newtonian_noise_reduction} including information about observed anisotropies of the seismic field. Anisotropy does not change our conclusion that NN is significantly reduced in Virgo due to the recesses.

\section{Summary and Outlook}

We have re-estimated the Newtonian noise considering the existence of clean rooms or recess-like structure underneath the test mirrors of the Virgo detector. Accountability of these recesses in numerical simulation leads to suppression of Newtonian noise relative to expectations mostly due to the fact that a recess increases the distance between test mass and ground. We obtain a significant suppression factor (by 2 at 15\,Hz) in the overall Newtonian-noise spectrum, which is important to include in future Newtonian-noise models for Virgo.  

We also investigated the impact of field anisotropy on the reduction factor. Especially at Virgo end buildings, knowledge of source locations (ventilation, pumps, ...) and seismic-array analyses already indicate an anisotropy of the field \cite{TrEA2019}. We found that the direction of propagation of the seismic field is important, but reduction of Newtonian noise is achieved for almost all propagation directions except for those where Newtonian noise without a recess would vanish, i.e., the practically irrelevant case where all Rayleigh waves propagate perpendicular to the detector arm.

The presence of recesses is effective especially for surface GW detectors where the dominant seismic Newtonian noise comes from comparatively slow Rayleigh waves ($\lesssim$300\,m/s). In this case, the required horizontal extent of the recess below test masses is of order 10\,m, which is a feasible modification of a detector infrastructure (as demonstrated at the Virgo site). Reduction of Rayleigh-wave Newtonian noise by a factor 2 and more is not minor given that a similar reduction by means of Newtonian-noise cancellation requires large arrays. Better suppression can be achieved by choosing optimized recess geometries as shown in past work \cite{HaHi2014}. However, given the high seismic speeds in underground environments, it seems unlikely that a similar mechanism can be exploited in the Einstein Telescope \cite{ET2011}, which would essentially ask for cavern sizes of around 100\,m to be effective.

\section*{Acknowledgements}

The authors thank Andrea Paoli for providing access to the engineering drawings of the Virgo buildings. The authors are grateful to colleagues from the Virgo collaboration and the LIGO Scientific Collaboration for fruitful discussions. A. S. is grateful for support via a Ron Drever Fellowship at the University of Glasgow.

\bibliographystyle{apsrev} 
\bibliography{references}

\end{document}